\documentclass[conference]{IEEEtran}
\IEEEoverridecommandlockouts
\usepackage{cite}
\usepackage{amsmath,amssymb,amsfonts}
\usepackage{algorithmic}
\usepackage{graphicx}
\usepackage{textcomp}
\usepackage{xcolor}
\usepackage{hyperref}
\usepackage{comment}
\def\BibTeX{{\rm B\kern-.05em{\sc i\kern-.025em b}\kern-.08em
    T\kern-.1667em\lower.7ex\hbox{E}\kern-.125emX}}
\begin{document}

\title{Fine-tuning Wav2vec for Vocal-burst Emotion Recognition}

\author{
\IEEEauthorblockN{Dang-Khanh Nguyen}
\IEEEauthorblockA{\textit{Department of AI Convergence} \\
\textit{Chonnam National University}\\
Gwangju, South Korea \\
nguyendangkhanh@jnu.ac.kr}
\and
\IEEEauthorblockN{Ngoc-Huynh Ho}
\IEEEauthorblockA{\textit{Department of AI Convergence} \\
\textit{Chonnam National University}\\
Gwangju, South Korea \\
nhho@jnu.ac.kr}
\and
\IEEEauthorblockN{Sudarshan Pant}
\IEEEauthorblockA{\textit{Department of AI Convergence} \\
\textit{Chonnam National University}\\
Gwangju, South Korea \\
sudarshan@jnu.ac.kr}
\and
\IEEEauthorblockN{Guee-Sang Lee}
\IEEEauthorblockA{\textit{Department of AI Convergence} \\
\textit{Chonnam National University}\\
Gwangju, South Korea \\
gslee@jnu.ac.kr}
\and
\IEEEauthorblockN{Soo-Huyng Kim}
\IEEEauthorblockA{\textit{Department of AI Convergence} \\
\textit{Chonnam National University}\\
Gwangju, South Korea \\
shkim@jnu.ac.kr}
\and
\IEEEauthorblockN{Hyung-Jeong Yang*}
\IEEEauthorblockA{\textit{Department of AI Convergence} \\
\textit{Chonnam National University}\\
Gwangju, South Korea \\
hjyang@jnu.ac.kr}
}
\maketitle

\begin{abstract}
The ACII Affective Vocal Bursts (A-VB) competition introduces a new topic in affective computing, which is understanding emotional expression using the non-verbal sound of humans. We are familiar with emotion recognition via verbal vocal or facial expression. However, the vocal bursts such as laughs, cries, and signs, are not exploited even though they are very informative for behavior analysis. The A-VB competition comprises four tasks that explore non-verbal information in different spaces. This technical report describes the method and the result of SclabCNU Team for the tasks of the challenge. We achieved promising results compared to the baseline model provided by the organizers.
\end{abstract}

\begin{IEEEkeywords}
Vocal Burst, Emotion Recognition, Wav2vec
\end{IEEEkeywords}

\section{Introduction}

Language and facial expression are strong indicators for behavior analysis. There is numerous research trying to solve the emotion recognition problem based on these data. However, the non-linguistic vocalizations are understudied even though they are very useful information. Analyzing and applying these signals are interesting topics and require more attention from researchers. The A-VB competition was conducted for that reason and is expected to explore advanced improvements in emotion science. The competition \cite{b2} consists of 4 individual tasks as below:

\begin{itemize}
\item High-dimension task (A-VB-HIGH): a multi-output regression task generating 10 values in the range of [0,1] corresponding to levels of Awe, Excitement, Amusement, Awkwardness, Fear, Horror, Distress, Triumph, Sadness, and Surprise.
\item Two-dimension task (A-VB-TWO): a multi-output regression task generating 2 values in the range of [0,1] corresponding to levels of Valence and Arousal.
\item Culture task (A-VB-CULTURE): a multi-output regression task generating 40 values in the range of [0,1] corresponding to levels of culture including China, US, South Africa, and Venezuela combined with  levels of emotion including Awe, Excitement, Amusement, Awkwardness, Fear, Horror, Distress, Triumph, Sadness, Surprise.
\item Type task (A-VB-TYPE): a multi-class classification task predicting the type of expressive vocal including Gasp, Laugh, Cry, Scream, Grunt, Groan, Pant, Other.
\end{itemize}

The evaluation metric for three regression tasks is the Concordance Correlation Coefficient (CCC) and the metric for the categorical task is the Unweighted Average Recall (UAR). The detail is listed below:
\begin{itemize}
\item A-VB-HIGH: The metric is the average CCC score of 10 emotions.
\item A-VB-TWO: The metric is the average CCC score of valence-arousal.
\item A-VB-CULTURE: The metric is the average CCC score of 40 culture-emotion levels.
\item A-VB-TYPE: The metric is the UAR score of 8 classes of vocalizations.
\end{itemize}

All of the above metrics are in the range of [0,1]. The greater the score is, the better the model performs. We should also note that all the results in this paper are in percentage. 

In this paper, we propose a straightforward approach using a pre-trained Wav2vec network to resolve the problem. The model accomplishes a noticeable improvement compared to the baseline provided by the organizers. Because of its simplicity, our model can be considered a new baseline for all tasks in the competition.

\section{Related work}
In the baseline paper \cite{b2}, the authors introduce two different approaches, which are feature-based and end-to-end approaches. In the feature-based option, the OpenSMILE toolkit \cite{b11} is leveraged to extract the COMPARE (COMputational PARalinguistics ChallengE \cite{b12}) feature and EGEMAPS (The extended Geneva Minimalistic Acoustic Parameter Set \cite{b13}) feature from an input sample. The features are then fed to a 3-layer fully-connected neural network. Mean Squared Error (MSE) loss is used for regression tasks while the classification task applies the Cross-entropy (CE) loss function.

In the end-to-end manner, the baseline model uses End2You \cite{b14}, a multimodal profiling toolkit that is capable of training and evaluating models from raw input. Particularly, Emo-18 architecture \cite{b15} is chosen for the competition. The model includes 1-D Convolutional Neural Network (CNN) layers to derive the features from audio frames and a Recurrent Neural Network to learn the temporal information.

For the speaker recognition task, Nik Vaessen and David A. van Leeuwen \cite{b4} conducted fine-tuning the Wav2vec2 model by using a shared fully-connected layer. Their model and ours have one thing in common: exploiting the pre-trained Wav2vec. However, there are considerable differences between the two methods. Basically, speaker recognition is a classification problem so the authors optimize the model with CE or Binary Cross-entropy (BCE) loss. In our method, we consider two loss options for the regression tasks, which are MSE and CCC loss. Additionally, besides using a shared fully-connected layer, we also take advantage of the RNN to explore the temporal behavior.
\section{Method}
\begin{figure}[htbp]
\centerline{\includegraphics[height=8cm]{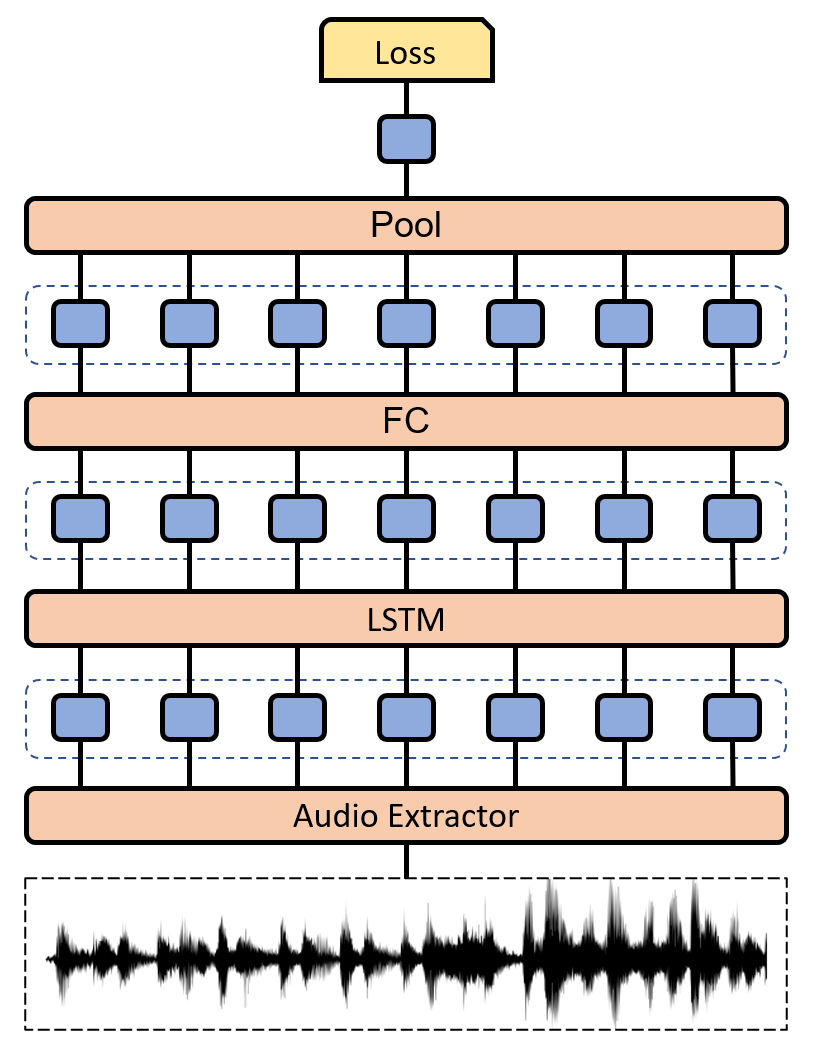}}
\caption{Block diagram of our proposed model.}
\label{fig:model}
\end{figure}

The sequence embeddings are obtained from the waveform signal by the audio extractor. They are then fed into an RNN to enrich the sequence information. Afterward, a fully connected layer changes the embeddings' dimension to the output sizes depending on the particular task. Finally, a pooling layer is used to reduce variable-length sequence embeddings to fix-size speaker embedding. The dimension of the final prediction would be 2, 10, 40, or 8 corresponding to A-VB-TWO, A-VB-HIGH, A-VB-CULTURE or A-VB-TYPE task, respectively. Figure~\ref{fig:model} describes the architecture of our method.

\subsection{Audio extractor}

We take advantage of the Pre-trained Wav2vec 2.0 models \cite{b3} provided by Pytorch. They are trained with large unlabeled audio corpora so they can effectively capture the audio features. We conducted experiments with 4 versions of the Wav2vec2 model described below:
\begin{itemize}
\item BASE: use the base configuration of transformer trained with 960 hours of unlabeled audio from LibriSpeech \cite{b5}.
\item LARGE: use the large configuration of transformer trained with 960 hours of unlabeled audio from LibriSpeech \cite{b5}.
\item LARGE-LV60K: use the large configuration of transformer trained with 60,000 hours of unlabeled audio from Libri-Light \cite{b6}.
\item XLSR53: use base configuration of transformer \cite{b10} trained with 56,000 hours of unlabeled audio from multiple datasets (Multilingual LibriSpeech  \cite{b7}, CommonVoice  \cite{b8} and BABEL  \cite{b9}).
\end{itemize}

\subsection{Pooling Method}
Inspired by the model of \cite{b4}, we use 4 options of pooling to fix the length of embedding: first (take the first sequence embedding to be the speaker embedding), last (take the last sequence embedding to be the speaker embedding), max, and average pooling. The performance of models with various types of pooling layers is recorded to study their impact on the result. The operation of the pooling can be described as:
\begin{equation}
s_{i} = Pooling(e_1, e_2, ..., e_{m_i})
\end{equation}
where $s_{i}$ is the speaker embedding of $i^{th}$ sample; $e_1, e_2, ..., e_{m_i}$ are the temporal embeddings and $m_i$ is the sequence length of corresponding sample.

\subsection{Loss function}
We use the CE loss for the classification task. In the remaining tasks, we want to test the effect of the loss function on the performance of the model so we did the experiments and compared the result of the model using Mean Square Error (MSE) and Concordance Correlation Coefficient (CCC) loss. The CCC loss is formulated as below:
\begin{equation}
\mathcal{L} = 1 - CCC = 1 - \frac{2s_{xy}}{s^{2}_x + s^{2}_y  + \left(\overline{x}-\overline{y}\right)^2}
\end{equation}
where $\overline{x}$ and $\overline{y}$ are the mean values of ground truth and predicted values, respectively, $s_x$ and $s_x$ are corresponding variances and $s_{xy}$ is the covariance value.

\section{Dataset and Experiments}
\subsection{Dataset}
The database used for the competition is the HUME-VB dataset  \cite{b1} which consists of 59201 audio files and is split into 3 sets (training, validation, and test) of similar size. Each file has 53 labels corresponding to 4 tasks, one label is a categorical label that is used in the classification task and the remains are values in the range [0,1] representing the emotional level. The organizers provide 2 versions of the HUME-VB dataset: the raw version sampled at 48kHz and the processed version where audio files are converted to 16kHz and normalized to -3 decibels. In our experiments, we take the processed version as the input of our model.

\subsection{Experiments}
Our model was implemented using the Pytorch framework. The experiments were conducted on a machine with NVIDIA RTX 2080 Ti GPU. All scenarios were run in 20 epochs, and the model with the best performance on the validation set was recorded. The batch size is 16 and the initial learning rate is $1e-4$. We used Adam optimizer with the weight decay coefficient of $0.0625$.

In our setting, we take the output from the $12^{th}$ layer of the Wav2vec network to be the sequence embeddings. The input size of the RNN network depends on the configuration of the pre-trained audio extractor, which is 768 for base configuration and 1024 for large configuration. It includes 2 LSTM layers and the hidden size is fixed to 512.

\section{Discussion}
We tested the performance of the model with various audio extractors to explore their effect. Table~\ref{tab:result} shows the performance on four tasks of the competition with 4 pre-trained Wav2vec extractors. As a result, the XLSR53 pre-trained model achieves the best performance in A-VB-TWO and A-VB-TYPE when LARGE-LV60K attains the highest scores in A-VB-HIGH and A-VB-CULTURE. In the meanwhile, the BASE model produces the lowest score in A-VB-TWO and A-VB-TYPE due to its simple architecture.

\begin{table}
 \caption{Evaluation score on the HUME-VB validation set with different extractors. Experimented with CCC loss and Last Pooling}
\begin{center}
  \centering
  \begin{tabular}{|l|l|l|l|l|l|}
    \hline
    Model & TWO & HIGH & CULTURE & TYPE \\
    \hline
    \textit{End2You Baseline} & \textit{49.88}  & \textit{56.38} & \textit{43.59} & \textit{41.66} \\
    \hline
    BASE & 54.65  & 58.69 & 47.18 & 41.65     \\
    \hline
    LARGE & 55.42 & 58.00 & 47.12 & 43.96      \\
    \hline
    LARGE-LV60K & 61.59 & \textbf{65.41} & \textbf{53.39} & 47.22  \\
    \hline
    XLSR53 & \textbf{61.94} & 65.32 & 52.50 & \textbf{49.89}  \\
    \hline
  \end{tabular}
  \label{tab:result}
\end{center}
\end{table}

Regarding the pooling method, we examined their influence on the results in A-VB-HIGH. As shown in Table~\ref{tab:pool}, in both LARGE-LV60K and XLSR53 model, the Last pooling outperforms the other options while the First pooling gets the lowest CCC score among the four methods. The result of Avg pooling is slightly better than Max pooling in both LARGE-LV60K and XLSR53 scenarios. It can be inferred that the last embedding of the sequence contains the most useful information for the prediction when using other embeddings or combining them may degrade the accuracy.

\begin{table}
 \caption{Evaluation score on the HUME-VB validation set with different pooling methods. Experimented on A-VB-HIGH with CCC loss}
\begin{center}
  \centering
  \begin{tabular}{|l|l|l|}
    \hline
    Pooling & LARGE-LV60K & XLSR53 \\
    \hline
     First & 53.56 & 58.20 \\
    \hline
     Max & 60.08 & 61.41 \\
    \hline
     Avg & 61.49 & 62.40 \\
    \hline
     Last & \textbf{65.41} & \textbf{65.32} \\
    \hline
  \end{tabular}
  \label{tab:pool}
\end{center}
\end{table}
Next, we conducted the training processes with MSE and CCC to explore their advantage. As a consequence, the model trained with CCC loss gives better performance on the validation set compared to the one trained with MSE loss. The detail is shown in Table~\ref{tab:loss}.
\begin{table}
 \caption{Evaluation score on the HUME-VB validation set with different loss functions. Experimented on A-VB-HIGH with Last pooling}
\begin{center}
  \centering
  \begin{tabular}{|l|l|l|}
  \hline
    Loss function & LARGE-LV60K & XLSR53 \\
    \hline
    MSE & 63.77 & 63.94 \\
    \hline
    CCC & \textbf{65.41} & \textbf{65.32} \\
    \hline
  \end{tabular}
  \label{tab:loss}
\end{center}
\end{table}

In addition, we carried out the ablation study to analyze the role of the RNN. According to Table~\ref{tab:rnn}, using the RNN can significantly boost the accuracy of the model in all four tasks. It can be explained by the capability of learning temporal information of the LSTM, which can enhance the overall operation of the model.

\begin{table}
 \caption{Evaluation score on the HUME-VB validation set with and without RNN. Experimented on A-VB-HIGH with LARGE-LV60K model, CCC loss, and Last pooling}
\begin{center}
  \begin{tabular}{|l|l|l|l|l|}
    \hline
    Model & TWO & HIGH & CULTURE & TYPE \\
    \hline
    Without RNN & 47.38 & 50.70 & 40.20 & 39.10 \\
    \hline
    With RNN & \textbf{61.59} & \textbf{65.41} & \textbf{53.39} & \textbf{47.22} \\
    \hline
  \end{tabular}
  \label{tab:rnn}
\end{center}
\end{table}

After conducting the above experiments, we concluded that the best configuration of our model is combining either LARGE-LV60K or XLSR53 pre-trained model with last pooling method and utilizing CCC loss. This setting was used to train separated model for each task in order to obtain unbiased evaluation on test set. We decided to choose LARGE-LV60K for A-VB-HIGH and A-VB-CULTURE, XLSR53 for A-VB-TWO and A-VB-TYPE. This time we trained each model for 50 epochs and applied early stopping by monitoring the validation result. Our best models and their evaluations on test set and validation set are listed in  Table~\ref{tab:test}.

\begin{table}
\caption{Evaluation score on the HUME-VB validation and test sets. Experimented with CCC loss and Last pooling for 50 epochs}
\begin{center}
\begin{tabular}{|c|c|c|c|}
\hline
 Task & Pre-trained &\multicolumn{2}{|c|}{Performance} \\
\cline{3-4} 
 name & Audio Extractor & Val set & Test set \\
\hline
 TWO & XLSR53 & 61.94 & 62.02 \\
\hline
 HIGH & LARGE-LV60K & 66.76 & 66.77 \\
\hline
 CULTURE & LARGE-LV60K & 54.93 & 54.95 \\
\hline
 TYPE & XLSR53 & 49.89 & 49.70 \\
\hline
\end{tabular}
\label{tab:test}
\end{center}
\end{table}

\section{Conclusion}
This paper presents our proposed method for all sub-challenges of the A-VB competition. Particularly, we fine-tuned the pre-trained Wav2vec and combined it with basic neural networks and a proper pooling method. The CCC loss and Last pooling show the best performance on four tasks among the other options. Our model outperforms the baseline of the organizer on the test set, with the CCC score of 62.02 for A-VB-TWO, 66.77 for A-VB-HIGH, 54.95 for A-VB-CULTURE and UAR metric of 49.70 for A-VB-TYPE.

\section*{Acknowledgment}

This work was supported by a National Research Foundation of Korea (NRF) grant funded by the Korean government (MSIT). (NRF-2020R1A4A1019191).



\begin{thebibliography}{00}
\bibitem{b1} Cowen, Alan and Baird, Alice and Tzirakis, Panagiotis and Opara, Michael and Kim, Lauren and Brooks, Jeff and Metrick, Jacob, ``The hume vocal burst competition dataset (H-VB) | raw data [exvo: updated 02.28.22] [data set],'' Zenodo, 2022.
\bibitem{b2} Baird, Alice and Tzirakis, Panagiotis and Batliner, Anton and  Schuller, Björn and Keltner, Dacher and Cowen, Alan, ``The ACII 2022 Affective Vocal Bursts Workshop \& Competition: Understanding a critically understudied modality of emotional expression,'' arXiv preprint arXiv:2207.03572 (2022).
\bibitem{b3} Baevski, Alexei, Yuhao Zhou, Abdelrahman Mohamed, and Michael Auli, ``wav2vec 2.0: A framework for self-supervised learning of speech representations,'' Advances in Neural Information Processing Systems 33 (2020): 12449-12460.
\bibitem{b4} Vaessen, Nik, and David A. Van Leeuwen, ``Fine-tuning wav2vec2 for speaker recognition,'' In ICASSP 2022-2022 IEEE International Conference on Acoustics, Speech and Signal Processing (ICASSP), pp. 7967-7971. IEEE, 2022.
\bibitem{b5} Vassil Panayotov, Guoguo Chen, Daniel Povey, and Sanjeev Khudanpur, ``Librispeech: an asr corpus based on public domain audio books,'' In 2015 IEEE International Conference on Acoustics, Speech and Signal Processing (ICASSP), volume, 5206–5210. 2015.
\bibitem{b6} J. Kahn, M. Rivière, W. Zheng, E. Kharitonov, Q. Xu, P. E. Mazaré, J. Karadayi, V. Liptchinsky, R. Collobert, C. Fuegen, T. Likhomanenko, G. Synnaeve, A. Joulin, A. Mohamed, and E. Dupoux, ``Libri-light: a benchmark for asr with limited or no supervision,'' In ICASSP 2020 - 2020 IEEE International Conference on Acoustics, Speech and Signal Processing (ICASSP), 7669–7673. 2020.
\bibitem{b7} Vineel Pratap, Qiantong Xu, Anuroop Sriram, Gabriel Synnaeve, and Ronan Collobert, ``Mls: a large-scale multilingual dataset for speech research,'' Interspeech 2020, Oct 2020.
\bibitem{b8} Rosana Ardila, Megan Branson, Kelly Davis, Michael Henretty, Michael Kohler, Josh Meyer, Reuben Morais, Lindsay Saunders, Francis M. Tyers, and Gregor Weber, ``Common voice: a massively-multilingual speech corpus,'' 2020.
\bibitem{b9} Mark John Francis Gales, Kate Knill, Anton Ragni, and Shakti Prasad Rath, ``Speech recognition and keyword spotting for low-resource languages: babel project research at cued,'' In SLTU. 2014.
\bibitem{b10} Alexis Conneau, Alexei Baevski, Ronan Collobert, Abdelrahman Mohamed, and Michael Auli, ``Unsupervised cross-lingual representation learning for speech recognition,'' 2020.
\bibitem{b11} F. Eyben, M. Wöllmer, and B. Schuller, ``Opensmile: the Munich versatile and fast open-source audio feature extractor,'' in Proceedings of the 18th ACM International Conference on Multimedia, 2010, pp.
1459–1462.
\bibitem{b12} B. Schuller, S. Steidl, A. Batliner, J. Hirschberg, J. K. Burgoon, A. Baird, A. Elkins, Y. Zhang, E. Coutinho, K. Evanini et al., ``The interspeech 2016 computational paralinguistics challenge: Deception, sincerity \& native language,'' in Proceedings of INTERSPEECH, 2016, pp. 2001–2005.
\bibitem{b13} F. Eyben, K. R. Scherer, B. W. Schuller, J. Sundberg, E. André, C. Busso, L. Y. Devillers, J. Epps, P. Laukka, S. S. Narayanan et al., ``The geneva minimalistic acoustic parameter set (gemaps) for voice research and affective computing,'' IEEE Transactions on Affective Computing, vol. 7, no. 2, pp. 190–202, 2015.
\bibitem{b14} P. Tzirakis, S. Zafeiriou, and B. W. Schuller, ``End2you–the imperial toolkit for multimodal profiling by end-to-end learning,'' arXiv preprint arXiv:1802.01115, 2018.
\bibitem{b15} P. Tzirakis, J. Zhang, and B. W. Schuller, ``End-to-end speech emotion recognition using deep neural networks,'' In Proc. ICASSP, 2018, pp. 5089-5093.
\end{thebibliography}
\end{document}